\documentstyle[epsf,twocolumn,aps,prb]{revtex}
\begin{document}
\draft
\twocolumn[\hsize\textwidth\columnwidth\hsize\csname @twocolumnfalse\endcsname
\title{Pairing Theory of High and Low Temperature Superconductors}
\author{Sang Boo NAM}
\address{University Research Center, Wright State University, 7735 Peters Pike, Dayton, OH 45414, USA \\
Physics  Department, POSTECH, Pohang 790-784, Korea \\
 Invited talk at ICAMD (1996)}
\maketitle
\vspace*{3mm}

~~~A scenario which can account for all observed features of both high-$T_c$ superconductors (HTS) and low-$T_c$ 
superconductors (LTS) is discussed. 
This scenario is based on the fact that a finite pairing interaction energy range 
$T_d$ is required to have a finite  value of $T_c$ and that not  all carriers participate in pairings, 
yielding multiconnected superconductors  (MS). A  new density  of states,  derived by keeping  the 
order parameter zero outside  of $T_d$, is  shown to account  for the observed low energy states in HTS and for the temperature dependences in the specific heat, the penetration depth, the optical conductivity, 
 and the tunneling conductance data.
I argue that the notion of MS can account for the tunneling data along the a(or b)$-$, ab$-$, and c$-$axis, 
and the 1/2 flux  quantum observed in   HTS. The region  occupied by  unpaired carriers  can be 
considered as a vortex  with a fluxoid quantum  number equal to  1 (VF), 0 (VZF), or  $-1$ (VAF) 
when the magnetic flux around the vortex is greater than, equal to, or less than the effective 
flux produced by the supercurrent, respectively.
The Hall anomaly depends on the relative strengths of the contributions via VF and VAF. The fact that 
the present scenario  can account  for all  observed features of  HTS and  LTS suggests  that the 
symmetry of the order parameter in HTS may not be different from one in LTS.
\vspace*{5mm}
]

\section{INTRODUCTION}
Superconductivity is one of the macroscopic quantum phenomena in nature.
Understanding the  origin of  this phenomenon  is one  of the challenging  problems[1] in  condensed 
matter physics.
In the   low-$T_c$ superconductors   (LTS), superconductivity  has been   well 
understood by Bardeen, Cooper,  and Schrieffers  (BCS) pairing  theory[2,3]. However,  in 
high-$T_c$ superconductors (HTS)[4, 5], the pairing mechanism remains unclear.  Various proposals for HTS have been made, but there is no clear consensus on the correct mechanism.  All of these approaches are based on pairing of carriers. Here, I discuss superconductivity 
in HTS and LTS based on the  fact that pairs are formed within a  finite interaction energy range 
$T_d$.
The fluxoid   quantization in   HTS[6] indicates   that pairing  of  carriers  are  responsible  for 
superconductivity. The  isotope effect  measured as  a function  of  $T_c$ suggests[7]  that the 
electron-phonon interaction may play an important role in HTS.
However, the   BCS results   are incompatible  with  the  experiments  which probe   low-energy 
excitations. For example, the temperature dependence of the specific  heat[8,9], the penetration depth 
[10-12], the  conductance  in the   normal superconducting  tunnel junction[13],   and the optical 
conductivity[14] can not be understood by using the BCS density of states.  To understand HTS 
within the framework of the pairing theory, various pairing states have been suggested[15]. The 
symmetry of the order parameter (i.e.,  $s-$ or $d-$wave symmetry)  is one of the issues  which 
still  remain  controversial.   The data   on  interference  associated   with two   weakly  coupled 
superconductors made  with YBCO  epitaxial films[16],  YBCO$-Pb$ tunnel  junction along  the 
$c-$axis direction[17] and microwave induced steps[18], suggest the $s$-wave is the correct symmetry for the order parameter.  However, the data on 
YBCO$-Pb$ SQUID's  and on tunnel  junctions along  the $ab-$ direction[19]  and the observed  half flux 
quantum in a YBCO ring[20] indicate the $d$-wave symmetry.

In this paper, I review  a scenario[21] which can  account for all observed  features of HTS as 
well as LTS. The  main idea is  based on the fact  that within the  pairing theory a finite  pairing 
interaction energy range $T_d$  is needed  to have a  finite value  of $T_c$ because  $T_c$ scales  with 
$T_d$. In other words, the pairs  are formed only within $T_d$,  and the order parameter may  be 
written as
\begin{eqnarray}
\Delta (k,\omega) = \left\{\begin{array}{l} \Delta ,~{\rm for}~|\epsilon_k| < T_d\\
0, ~~{\rm for}~|\epsilon_k|> T_d\end{array}\right.  
\label{eq1}
\end{eqnarray}
Here, $\epsilon_k$ is the normal-state  excitation energy of momentum $k$  measured with respect 
to the  Fermi  energy. The  frequency  $\omega$ has   no restriction. Equation   (1) can  be also 
applicable to describing the electron-phonon  interaction within the BCS  theory because the pairing 
interaction is attractive near the  Fermi level. I point  out that consistent treatment  of Eq. (1) can 
lead to a new physics.  When the pairs are  formed, low energy states  are pushed to high energy 
states which are above $\Delta$. Conversely, when the pairs are broken for $|\epsilon| >T_d$, the electrons return to the states where they came from. 
This leads to a new density of states [see Eq. (4) below] $q(\omega/T_d)$. I argue that this can account for low energy states observed in  HTS. For  a large  value of $y=T_d/\pi  T_c$ (i.e.,  LTS), $q(\omega/T_d)$  is 
negligible, and the BCS result is recovered. Therefore, all  the calculations based on the BCS results 
are reliable only for a large  value of $y$ (i.e., $y \gg  1$). However, in HTS, the values of $y$ is 
not large. Hence, it is essential to obtain a solution  which is valid for all values of $y$ in order  to 
test the pairing theory for HTS.

\section{TRANSITION TEMPERATURE}

The pairing   mechanism is  unclear, but   the isotope  effect  suggests that and electron-phonon 
interaction may play an important role in HTS. Here, I consider the electron-phonon type mechanism 
and obtain an exact equation for $T_c$ by using the BCS equation as[21]
\begin{eqnarray}
f(y) =   \frac{2}{\pi} \displaystyle{\sum_j}\frac{2}{j}\tan^{-1}\frac{y}{j}   = \frac{1}{g} = \int^y_0  \frac{dx}{x} 
\tan{\rm h}\frac{\pi x}{2}.
\end{eqnarray}\label{eq2}
\hspace*{-0.1cm}Here, the $j$   are odd  integers,   $g$  corresponds  to   the  BCS  pairing   interaction  parameter 
$N(0)V_{BCS}$, and $N(0)$ is the density of  states at the Fermi level. Because of  the arctangent 
function in Eq. (2), the sum converges. For  a large value of $y$,  
Eq. (2) yields the BCS result
\begin{eqnarray}
f_{BCS}(y) &=& \ln \frac{y}{0.28} = \frac{1}{g}\nonumber ,\\
T_c(BCS) &=& 1.14T_d \exp\left(-\frac{1}{g}\right). \label{eq3} 
\end{eqnarray}
The BCS result breaks down when $y < 0.28$ because  $f_{BCS}(y) < 0$. For a given value of  $T_c, 
f(y)$ needs a  smaller value  of $g$  than $f_{BCS}(y)$.  For $f(0.28)=1/2.32, T_c  = 1.14T_d$  is 
obtained for $g = 2.32$. This is in contrast to $g = \infty$ for $f_{BCS}(y) = 0$. For $g > 2.32, f(y) = \pi y/2 = 1/g$ yields $T_c =  g T_d/2$. For $f(4/\pi)= 1/0.657, T_c = 
T_d/4 = 100~K$  with $T_d  = 400K$[22], is  obtained for  $g = 0.657$.  These values  may be 
physically realized in YBCO with the electron-phonon  coupling constant[23] $\lambda \sim 1.3$ 
to 2.3 because $g$ may be approximated  as $\lambda$ or $\lambda/(1+\lambda)$. In Fig. 1, $T_c$ 
of Eq.(2), from the present scenario, is compared with the BCS result of Eq. (3).

\begin{figure}[bht]
\centering
\epsfxsize=6cm
\epsffile{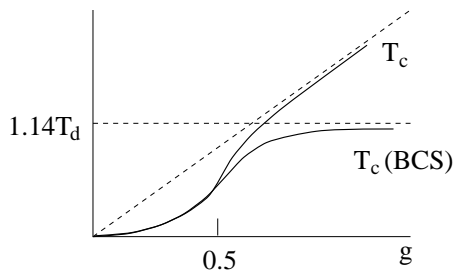}
\caption{Schematic diagram for the transition temperature.}
\end{figure}

The isotope effect[24], which is obtained  by taking the derivative of Eq. (2), is compatible with the  
data[7]. Moreover,  by using   a generalized expression   of $g$, which   includes the Coulomb 
interaction, the dynamical screening factor (DSF) in the Coulomb interaction can be shown to play an 
important role in the doping concentration dependence of  $T_c$[24]. DSF decreases with increasing 
$T_c$[25]. When $T_c$ increases initially with doping, DSF decreases and makes the repulsive 
interaction increase. Thus,  $T_c$ is expected  to have  a maximum at  the optimum doping  as a 
function of the doping concentration[26].

\section{NEW DENSITY OF STATES}

During the last four decades, the BCS density of states 
$n_{BCS}(\omega)=Re[\omega/(\omega^2-\Delta^2)^{1/2}]$ has been used in understanding most of the 
data in LTS. However,  $n_{BCS}(\omega)$ is not compatible with low energy states observed in HTS. I 
present below that low energy states arise from the pair breaking processes for $|\epsilon_k|>T_d$.

The density of states is obtained by perfoming the  $\epsilon_k$ integration of the imaginary part of the Green function[3]. 
By carrying out the $\epsilon_k$ integration with the condition that Eq. (1) 
is satisfied, we obtain the density of states as[21, 27]
\begin{eqnarray}
\frac{N(\omega)}{N(0)} = n(\omega)=q(\omega/T_d)+n_{BCS}(\omega)r(\omega/T_d),&&\\\label{eq4}
q(\omega/T_d)=\frac{2}{\pi} \tan^{-1}\frac{\omega}{T_d},&&\nonumber\\
r(\omega/T_d)=\frac{2}{\pi}\tan^{-1}\frac{n_{BCS}(\omega)T_d}{\omega}.\nonumber
\end{eqnarray}
In the large $T_d$ limit, the $q(\omega/T_d)$ term vanishes and $r(\omega/T_d)\rightarrow 1$. Hence, $n(\omega) \rightarrow n_{BCS}(\omega)$. Because $q(\omega/T_d)$ vanishes at zero frequency, a superconductor has a finite order parameter $\Delta$, but, in principle, there is no excitation energy gap. In LTS, $q(\omega/T_d)$ is negligible so that the excitation energy gap observed is $\Delta$. I note that $q(\omega/T_d)$ remains the same even when the retardation effect is taken into account. The retardation effect can be included in $\Delta$ by writting $\Delta(\omega)$ in Eq. (1). The usual picture of the density of states for quasiparticles $d\epsilon_k/dE$ is no longer useful because $\epsilon_k$ is not a single valued function of $E$. In Fig. 2, the new density of states $n(\omega)$ is compared with $n_{BCS}(\omega)$.

\begin{figure}[bht]
\centering
\epsfxsize=6cm
\epsffile{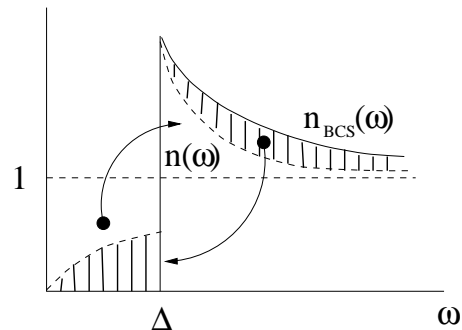}
\caption{Schematic diagram representing the qualitative difference
 between $n(\omega)$ and $n_{\scriptscriptstyle BCS}^{}(\omega)$.}
\end{figure}

By using $N(\omega)$, it is straightforward to calculate various properties[27]. I obtain the $T^2$ term in the thermally excited quasiparticle density at low $T$. This is compatible with the magnetic penetration depth data[10, 12], the $T^2$ dependence in the specific heat data in HTS[8,9] and heavy fermion superconductor (HFS)[28], and the $T^3$ dependence in the nuclear spin relaxation rate data in HTS[29] and HFS[30]. Moreover, $T$ and $T^2$ dependences are predicted for the conductivity. This behavior is consistent with the HTS data[14]. The conductance in the normal superconducting tunnel junction is predicted to be linear in $T$ for the zero-bias case and linear in the voltage at $T=0$. Again, this temperature dependence is consistent with the HTS data[13]. These $T$ dependences derived from $q(\omega/T_d)$ are due to the fact that not all carriers participate in pairings.
 
I consider the sum rule. The integration over $\omega$ for $n(\omega) - n_{BCS}(\omega)$ should vanish. Hence, I write $R(\Delta/T_d)$, which corresponds to the fraction of states not participating in pairings, as
\begin{eqnarray}
R(\Delta/T_d)\Delta&=& \int^\Delta_0 d\omega q(\omega/T_d)\nonumber\\\label{eq5}
&=& \Delta\left[q(\Delta/T_d)-\frac{T_d}{\pi\Delta}\ln\left(1+\frac{\Delta^2}{T^2_d}\right)\right]\nonumber\\
&=& \int^\infty_\Delta d\omega[n_{BCS}(\omega)-n(\omega)]\\\nonumber
\Delta&=& \int^\infty_\Delta d\omega[n_{BCS}(\omega)-1].\nonumber
\end{eqnarray}
For $T_c=T_d/4, R(0.458) = 14\%$. The condensed carrier density at $T=0$ may be written[21] as
\begin{eqnarray}
\frac{\rho_s}{\rho} = 1-R(\Delta/T_d).\label{eq6}
\end{eqnarray}
It is straightforward to see that the penetration depth would be longer than the London length by the factor of $[1-R(\Delta/T_d)]^{-1/2}$. Thus, complete condensation is not possible within the pairing theory because a finite value of $T_d$ is required for a finite value of $T_c$.

\section{MULTICONNECTED SUPERCONDUCTOR}

The incomplete condensation suggests the possibility of having multiconnected superconductors (MS). When the unpaired carriers are uniformly distributed, then a superconductor may be considered as simply connected.
In reality, this may be unlikely in HTS. For simplicity, the Fermi surface[31] of YBCO may be considered as the skeleton beams of a rectangular building with no wall and no floors.
The pairs are formed within $T_d$, and the order parameter, which is finite near the Fermi level, is anisotropic.
When an external magnetic field or current is applied to the system, then the flow patterns of the current in a two-dimensional view may be similar to the skeleton beams and yield rectangular fishing nets.
If the tunnel current flows in either the $a-$, $b-$, or $c-$direction, then no circulating current is possible.
Thus, the modulation of the Josephson current by magnetic fields yields an usual Fraunhofer pattern. However, when the tunnel current flows in the $ab-$direction, then there is a circulating current guarding the unpaired region from the small magnetic field inside the fishing net pattern.
This acts like a vortex with antiflux. Within this picture, one can have the 1/2 magnetic flux unit shift when the line integral of the current is taken along one or the other side around the unpaired region.
In other words, a circulating current between the two ends is the same as the reflections of the current at both ends.
Thus, a phase shift of $\pi$ (1/2 flux unit shift) occurs in the system of the odd number of MS sections.
This suggests that the notion[32] of a MS can account for the tunnel junction data[19] and the 1/2 flux quantum in HTS[20].

\section{VORTEX WITH ZERO FLUXOID}

Macroscopically, the current in a superconductor may be written $[\hbar=k_B=c=1$, and $\Phi_0 =\frac{h c}{2 e} =1]$ as
\begin{eqnarray}
J(r) = J_0(r)\left[\frac{\nabla\theta}{2\pi}-A(r)\frac{2e}{hc}\right]\label{eq7}
\end{eqnarray}
where $\theta$ is the phase of the order parameter, $A(r)$ is the vector potential, and $J_0(r)$ is proportional to the square of the order parameter amplitude times $2e/2m$. The London fluxiod quantum (L) condition may be obtained by the line integral of Eq. (7) around a contour as
\begin{eqnarray}
L = \oint dr\cdot \frac{1}{2\pi}\nabla\theta=\Phi_A + \Phi_J\label{eq8}
\end{eqnarray}
where $\Phi_A$ is the magnetic flux inside a contour and $\Phi_J$ is the effective flux via the line integral of $J(r)/J_0(r)$ along the contour.
$L$ is an integer. This reflects the single valuedness of the order parameter.

For simplicity, I consider a vortex (i.e., where the order parameter is zero) at the origin of the cylindrical coordinates $(r, \phi)$.
I omit the variation in the $z$-direction and write $\theta = L\phi$.
The phase term in Eq. (7) yields $L/r$. There are three possible cases for $L$ (i.e., $L = 1,0,$ and $-1$).
Therefore, the unpaired region in a MS may be considered as a vortex with a fluxoid quantum number equal to 1 (VF), 0 (VZF), or $-1$ (VAF) when the magnetic flux around a vortex is greater than, equal to, or less than the effective flux produced by the supercurrents, respectively.

In the case of VZF, the current goes with the vector potential (the London gauge).
VZF can not exist in a simply connected superconductor (SS). 
Because a vortex is formed after the magnetic flux has entered into a SS, only VF (i.e., L=1) is possible in a SS. 
When three different types of vortices are present in a superconductor, VZF breaks the global periodicity and acts as a domain wall between the VF and the VAF regions where local periodicities are maintained.
This can account for the regions of VF and VAF observed in Nb[33]. In other words, rather than the global free energy, the local free energy may be responsible for the formation of magnetic quantum states of vortices.

\section{ANOMALOUS HALL EFFECT}

It is well known that the motion of a vortex yields the Hall voltage in superconductors.
Vortices can[34] act like superconducting ringlets.
The Hall voltage produced by moving VF, VZF, and VAF can be easily obtained by considering them as stable noninteracting particles with magnetic quantum numbers equal to 1, 0, and $-1$, respectively.
The Hall angle $\theta_H$ may be written[34] as
\begin{eqnarray}
\tan\theta_H=\frac{e}{m}(n_1h_1-n_2h_2)\left(\frac{n_1}{\tau_1} + \frac{n_2}{\tau_2}\right)^{-1}\label{eq9}
\end{eqnarray}
where $n_1$ and $n_2$ are the VF and VAF number densities. $h_i$ depends on the force responsible for the motion of vortices: the magnetic induction B for the Lorentz force and the critical field $H_{c2}$ for the Magnus force.
$\tau_1$ and $\tau_2$ are the carrier scattering times in VF and VAF, respectively.
The sign of the Hall voltage depends on the relative strengthes of contributions from VF and VAF.
This is similar to a system with two oppositely charged carriers.
VZF does not contribute to Hall voltage because it moves in the direction of the transport current.
Hence, VZF does not yield an electric field or an energy dissipation.
The calculation[35] based on Eq. (9) suggests that this approach can describe the anomalous Hall effect data of both HTS and LTS.

\section{REMARKS ON OTHER MECHANISMS}

I make a few remarks on other mechanisms.
The recent angle resolved photoemission data[36] in $SrRuO_4~(T_c = 1~K)$ indicate that the van Hove singularity (vHS) is not sufficient condition for producing high $T_c$.
Moreover, the doping concentration dependence of the density of states in HTS and other systems[37] appear incompatible with the vHS scenario.
The anisotropic order parameter suggested via the angle resolved photoemission data[38] in HTS may only reflect the complexity of the Fermi surface. The observation of the quantum oscillations[39] in 
$SrRuO_4$ shows that the Fermi liquid picture may be valid and that the non-Fermi liquid scenario may not be necessary.
Schrieffer's[40] argument that the vertex correction reduces the strength of a nearly antiferromagnetic Fermi liquid (NAFL) questions the validity of the NAFL mechanism for HTS. The similarity between the magnetic field dependent specific heat data in $V_3 Si$ (LTS)[41] and YBCO[9] indicates that exotic pairing states may not be necessary to understand the data.

\section{CONCLUSIONS}

The fact that consistent solutions based on a finite pairing interaction energy range can account for all observed features of HTS and LTS, indicates that the symmetry of the
order parameter in HTS may not be different from the symmetry in LTS. This suggests that an exotic pairing mechanism may not be necessary to understand the origin of
superconductivity in HTS. I believe that the present theory will also be able to account for other unsolved problems in HTS.\\
 
I would like to thank J. H. Kim for many helpful discussions.


\begin{references}
\bibitem{a} The discoveries of HTS[4,5] have been drawing great attention in scientific as well  as social communities.
\end{references}
\end{document}